\documentclass[twocolumn,twocolappendix,tighten]{aastex631}

\usepackage[utf8]{inputenc}
\usepackage{xspace}
\usepackage{acronym}
\usepackage{amssymb}
\usepackage{pifont}
\usepackage{units}
\usepackage{comment}

\newcommand{\cmark}{\ding{51}}%
\newcommand{\xmark}{\ding{55}}%

\newcommand{\secref}[1]{{Sec.~(\ref{sec:#1})}}
\newcommand{\figref}[1]{{Fig.~(\ref{fig:#1})}}
\newcommand{\eqref}[1]{Eq.~(\ref{eq:#1})}
\newcommand{\tabref}[1]{{Table~\ref{tab:#1}}}
\newcommand{\appref}[1]{{Appendix~\ref{sec:#1}}}

\newcommand{\tn}[1]{\textnormal{#1}}
\newcommand{\tns}[1]{\scriptsize\textnormal{#1}}
\newcommand{\ye}[0]{Y_{\tns{e}}}

\newcommand{\fe}[0]{{}^{60}\tn{Fe}}
\newcommand{\pu}[0]{{}^{244}\tn{Pu}}
\newcommand{\mn}[0]{{}^{53}\tn{Mn}}
\newcommand{\rprocess}[0]{\textit{r}-process\@\xspace}

\newcommand{\msun}[0]{M_{\odot}}

\newcommand{\ie}[0]{i.e.\@\xspace}
\newcommand{\eg}[0]{e.g.\@\xspace}

\newacro{NSE}{nuclear statistical equilibrium}
\newacro{GCE}{galactic chemical evolution}
\newacro{EOS}{equation of state}
\newacroplural{EOS}{equations of state}
\newacro{BNS}{binary neutron star}
\newacro{NS}{neutron star}
\newacro{BH}{black hole}
\newacro{CC}{charged-current}
\newacro{CCSN}{core collapse supernova}
\newacro{GW}{gravitational wave}
\newacro{dof}{degrees of freedom}
\newacro{NR}{numerical relativity}
\newacro{LK}{leakage}
\newacro{LR}{low resolution}
\newacro{SR}{standard resolution}
\newacro{HR}{high resolution}
\newacro{RMF}{relativistic mean field}
\newacro{EM}{electromagnetic}
\newacro{AMR}{adaptive mesh refinement}
\newacro{ISM}{interstellar medium}
\newacro{SN}{supernova}
\newacro{Mya}{Myr ago}

\submitjournal{ApJL}

\received{October 24, 2023}
\revised{January 5, 2024}
\accepted{January 29, 2024}

\graphicspath{{./}}

\begin{document}

%\title{Template \aastex Article with Examples: 
%v6.3.1\footnote{Released on March, 1st, 2021}}

\title{Did a Kilonova Set Off in Our Galactic Backyard 3.5 Myr ago?}

\correspondingauthor{Leonardo Chiesa}
\email{leonardo.chiesa@unitn.it}

\author[0000-0002-5112-1975]{Leonardo Chiesa}
\affiliation{Dipartimento di Fisica, Università di Trento, via Sommarive 14,
  38123, Trento, Italy}
\affiliation{INFN-TIFPA, Trento Institute for Fundamental Physics and
  Applications, via Sommarive 14, 38123, Trento, Italy}

\author[0000-0002-0936-8237]{Albino Perego}
\affiliation{Dipartimento di Fisica, Università di Trento, via Sommarive 14,
  38123, Trento, Italy}
\affiliation{INFN-TIFPA, Trento Institute for Fundamental Physics and
  Applications, via Sommarive 14, 38123, Trento, Italy}
  
\author[0000-0003-3824-4433]{Federico Maria Guercilena}
\affiliation{INFN-TIFPA, Trento Institute for Fundamental Physics and
  Applications, via Sommarive 14, 38123, Trento, Italy}
\affiliation{Dipartimento di Fisica, Università di Trento, via Sommarive 14,
  38123, Trento, Italy}



\begin{abstract}
The recent detection of the live isotopes $\fe$ and $\pu$ in deep ocean
sediments dating back to the past \unit[3-4]{Myr} poses a serious challenge to
the identification of their production site(s). While $\fe$ is usually
attributed to standard core-collapse supernovae, actinides are
\rprocess nucleosynthesis yields, which are believed to be synthesized in rare
events, such as special classes of supernovae or binary mergers involving at
least one neutron star. Previous works concluded that a single binary neutron
star merger cannot explain the observed isotopic ratio. In this work, we
consider a set of numerical simulations of binary neutron star mergers producing
long-lived massive remnants expelling both dynamical and spiral-wave wind
ejecta. The latter, due to a stronger neutrino irradiation, produce also
iron-group elements. Assuming that large-scale mixing is inefficient before the
fading of the kilonova remnant and that the spiral-wave wind is sustained over a
\unit[100-200]{ms} timescale, the ejecta emitted at mid-high latitudes provide a
$\pu$ over $\fe$ ratio compatible with observations. The merger could have
happened \unit[80-150]{pc} away from the Earth, and between 3.5 and
\unit[4.5]{Myr} ago. We also compute expected isotopic ratios for eight other live
radioactive nuclides showing that the proposed binary neutron star merger
scenario is distinguishable from other scenarios proposed in the literature.
\end{abstract}


\section{Introduction}
\label{sec:intro}

The production of half of the elements heavier than iron in the Universe
(including all the elements heavier than lead) is due to the rapid neutron
capture process, \citep[$r$-process][]{Burbidge:1957vc}. Despite much progress
in the past few years, our precise understanding of the $r$-process and of its
yields is presently limited by uncertainties of both nuclear physical and
astrophysical nature \citep{Cowan:2019pkx}. The former are mostly due to the
paucity of experimental measurement of exotic neutron-rich nuclei, while the
latter are related to limitations in the modeling of the astrophysical sites.

Even the sites where $r$-process nucleosynthesis happens are still uncertain.
The detection of the kilonova AT2017gfo unambiguously associated to \ac{BNS}
merger GW170817 provided the first direct evidence of $r$-process
nucleosynthesis \citep{Pian:2017gtc,Smartt:2017fuw,Kasen:2017sxr}. The question
whether compact binary mergers involving at least one \ac{NS} are the only
relevant site is still debated. Compact binary mergers seem to have issues in
accounting for all available observables, including the abundances of
$r$-process elements in very metal-poor stars or in ultrafaint dwarf galaxies
\citep{Cote:2018qku,Bonetti:2019fxj}. Other possible $r$-process sites include
special types of (rare) \acp{SN}, such as magnetorotational supernovae
\citep{Winteler:2012hu,Mosta:2017geb} or collapsars \citep{Siegel:2018zxq},
although their viability is debated and they could be limited to low-metallicity
environments \citep[e.g.][]{Macias:2019oxw,Bartos:2019twj}.

The observation of $r$-process abundance patterns traceable to single events has
the potential to shed light on their production site. The detection of live (\ie
undecayed) radioactive isotopes in sediments is powerful in this respect, since
it features a nontrivial temporal dependence from their decay profiles
\citep{Ellis:1995qb,Wehmeyer:2023bmg}. There is a common consensus that the
$\fe$ (mean lifetime: $\tau_{\scriptsize\fe}$ \unit[$\simeq 3.8$]{Myr}) observed
in terrestrial and lunar samples points to one or more explosive events
happening $\lesssim 10$ \ac{Mya} not far from the Earth (\unit[$\lesssim
120$]{pc}). The production site is in general identified with a \ac{SN}, while
\ac{BNS} mergers were firmly disregarded
\citep{Fields:2004iq,Fry:2014yqa,Schulreich:2017dyt}. A concomitant finding
in deep-sea sediments of $\mn$ ($\tau_{\scriptsize\mn}$ \unit[$\simeq
5.4$]{Myr}), an isotope usually associated with Type Ia \ac{SN} events, has also
been reported at a similar depth of $\fe$ \citep{Korschinek:2020kfp}.

Recently, \citet{Wallner.etal:2016,Wallner2021} reported new measurements of
$\fe$ in deep-ocean sediments and ferromanganese crusts. The deduced
interstellar influx shows two peaks within the last \unit[10]{Myr}, the most
prominent one starting \unit[$\sim$3.5]{\ac{Mya}} and centered at
\unit[$\sim$2.5]{\ac{Mya}}, the smaller and narrower second one peaking at
\unit[$\sim$6.5]{\ac{Mya}}. Interestingly, they also documented the unambiguous
emergence of a $\pu$ ($\tau_{\scriptsize\pu}$ \unit[$\simeq 116.3$]{Myr})
signature, especially in association with the most recent and prominent $\fe$
peak. Based on their measurements, they reported an \ac{ISM} $\pu$ fluence at
Earth orbit of $\mathcal{F}_{\scriptsize\pu} = \unit[(7.7 \pm 1.6) \times
10^3]{atoms~cm^{-2}}$ and an abundance ratio of
$Y_{\scriptsize\pu}/Y_{\scriptsize\fe} = (53 \pm 6) \times 10^{-6}$ for the
\unit[0-4.6]{\ac{Mya}} time window. \citet{Wallner2021} attributed the two $\fe$
peaks to multiple nearby \acp{SN} happening within the last \unit[10]{Myr} at
50-\unit[100]{pc} from Earth. To explain the $\pu$ abundance, they considered a
pure \ac{SN} origin within the Local Bubble or a combination of \acp{SN} with a
previous nucleosynthesis event (e.g. a \ac{BNS} merger).

\citet{Wang2021} compared results from \citet{Wallner2021} with yield
predictions obtained from \ac{SN} and \ac{BNS} merger models. They claimed that
the observations are compatible with a single source located at \unit[$\lesssim
100$]{pc} only by considering a \ac{SN} with an enhanced \rprocess production,
while the scenario of a single nearby \ac{BNS} merger would be unfeasible. They
also proposed a two-step scenario, in which $\pu$ was produced by a rare and
more distant event polluting the Local Bubble, before reaching the Earth among
the debris from a nearby \ac{SN}. We notice that \citet{Wang2021} considered
isotropic ejecta. Moreover, the abundance yields of each source were obtained as
a linear combination of a few representative trajectories obtained in
simulations and fitted to reproduce metal-poor star observations.

In this \textit{Letter}, we revive the single \ac{BNS} merger scenario using
exclusively the outcome of \ac{BNS} merger simulations and the data reported by
\citet{Wallner2021}. We consider the case in which the merger produces a
long-lived remnant. By taking into account the prolonged effect of neutrino
irradiation and the resulting anisotropy in the nucleosynthesis yields, we find
that the coincident detection of $\fe$ and $\pu$ in the more recent portion of
the crust (\unit[$\lesssim 4$]{\ac{Mya}}) is compatible with a \ac{BNS} merger
event occurring at a distance of $\unit[\sim 80-150]{pc}$.

\section{Methods}
\label{sec:methods}

\begin{deluxetable}{ccccccc}
\tablecolumns{7}
\tablecaption{Summary of the Properties of the \ac{BNS} Merger Models Considered
in This Work}
\tablehead{
\colhead{\#} & \colhead{EOS} & \colhead{$q$} & \colhead{Vis} & \colhead{$t_{\rm end}$} & \colhead{$M_{\rm ej,dyn}$} & \colhead{$\dot{M}_{\rm ej,wind}$}  \\
\colhead{~} & \colhead{~} & \colhead{[-]} & \colhead{~} & (ms) & \colhead{($10^{-3}\msun$)} & \colhead{($10^{-1}\msun~{\rm s^{-1}}$)}
}
\startdata
 1 & BLh  & 1.0 & \cmark & 91.8 & 1.36 & 2.34 \\ %[1ex] 
 2 & BLh  & 0.7 & \cmark & 59.6 & 3.19 & 5.96 \\ %[1ex] 
 \hline
 3 & DD2  & 1.0 & \cmark & 113.0 & 1.47 & 1.97 \\ %[1ex] 
 4 & DD2*  & 0.83 & \xmark  & 91.0 & 2.25 & 1.79 \\ %[1ex] 
 \hline
5 & SFHo  & 0.7 & \cmark & 46.5 & 2.35 & 4.40 \\ %[1ex] 
\hline
 6 & SLy4  & 0.7 & \cmark & 40.3 & 1.98 & 4.72 \\ %[1ex] 
 \hline
\enddata 
\tablecomments{Each model is characterized by its \ac{EOS} and mass ratio $q$,
and was run until $t_{\tns{end}}$ postmerger, unbinding $M_{\tns{ej,dyn}}$ of
dynamical ejecta and spiral-wave wind ejecta with a rate
$\dot{M}_{\tns{ej,wind}}$. * denotes the only model without turbulent viscosity
(Vis).}
\label{tab:tab_1}
\end{deluxetable}

% introduce the simulations
We consider the outcome of six \ac{BNS} merger simulations originally presented in
\citet{Bernuzzi:2020txg,Nedora:2020pak} and part of the CORE database
\citep{Gonzalez:2022mgo}. They were targeted to GW170817 \citep[\ie their chirp
mass is 1.188~$\msun$, see][]{Abbott:2018wiz}, that we consider as
representative \ac{BNS} merger event and for which we explore different mass
ratios, $q \in \left[ 0.7,1 \right]$. A summary of the models is reported in
\tabref{tab_1}. Matter was evolved employing the \texttt{WhiskyTHC} code
\citep{Radice:2012cu, Radice:2013xpa, Radice:2013hxh}, complemented by a finite
temperature, composition-dependent \acf{EOS}. The \acp{EOS} used in those
simulations were BLh \citep{Logoteta:2020yxf}, HS(DD2) \citep[][hereafter
DD2]{Hempel:2009mc, Typel:2009sy}, SFHo \citep{Steiner:2012rk} and SRO(SLy4)
\citep[][hereafter SLy4]{Douchin:2001sv,daSilvaSchneider:2017jpg}.
% neutrinos in the simulations
Neutrino radiation was taken into account by a leakage scheme to model neutrino
emission and a M0 transport scheme to account for absorption in optically thin
conditions \citep{Radice:2016dwd,Radice:2018pdn}. These schemes were shown to
describe well the most relevant features of neutrino emission and reabsorption
\citep{Zappa:2022rpd}. The latter effect is crucial to predict the properties of
the unbound matter, ultimately influencing the ejecta composition
\citep[e.g.][]{Sekiguchi:2015dma,Foucart:2015gaa,Perego:2017wtu,Radice:2018pdn}.
% viscosity
In all but one simulation, turbulent viscosity of magnetic origin was included
via a large eddy scheme \citep{Radice:2017zta,Radice:2020ids}.
% resolution
Each simulation covered the innermost part of the domain with a grid of
resolution of $\Delta x = \unit[185]{m}$.

% characterize the simulations
In addition to unbind dynamical ejecta, these simulations produced a long-lived
merger remnant, lasting $t_{\tns{end}} \unit[\sim 40-110]{ms}$ postmerger and
showing no sign of gravitational collapse up to these times. Such a merger
outcome produces spiral-wave winds \citep{Nedora:2019jhl,Nedora:2020pak} that
possibly unbind an amount of matter significantly larger than the dynamical one.
Moreover, the longer exposition to neutrino irradiation increases the electron
fraction ($\ye$) in these ejecta.

% ejecta from the simulations
The unbound matter properties are extracted from the simulations on a sphere of
coordinate radius $R_{\tns{E}} = \unit[200]{\msun}\unit[\simeq295]{km}$. We use
the geodesic and the Bernoulli criterion to identify the dynamical and the
spiral-wave ejecta, respectively. We extract mass histograms in the space
characterized by the specific entropy ($s$), $\ye$ and expansion timescale
($\tau_{\tns{exp}}$). The latter is obtained from the radial speed and density
at $R_{\rm E}$ according to the method outlined in \citet{Radice:2016dwd,
Radice:2018pdn}. We keep track of the spatial composition of the ejecta
along the polar angle $\theta$ (measured with respect to the binary's orbital
axis), while we marginalize over the azimuthal angle.

% nucleosynthesis calculation
We compute the nucleosynthesis yields produced in each simulation convolving the
ejecta properties with the outcome of nuclear network calculations performed
with SkyNet \citep{Lippuner:2017tyn}. More details on how the isotopic
masses are extracted for each ejecta component are given in \appref{appendix}.
Overall, we compute the mass of each isotope $i$ at different polar angles as
the sum of the dynamical ($m_{{\rm ej},i}^{\rm dyn}$) and spiral-wave wind
($m_{{\rm ej},i}^{\rm wind}$) ejecta contribution at 30 years after merger.
Since the latter did not saturate at the end of the simulations, but had an
approximately constant ejection rate, we rescale the spiral-wave wind ejecta
yields by a factor $f_{\tns{wind}}(t_{\tns{wind}})
\equiv (t_{\tns{wind}} - t_{\tns{wstart}})/(t_{\tns{end}}-t_{\tns{wstart}})$,
such that $m_{\tns{ej},i} (\theta,t_{\tns{wind}})= m_{\tns{ej},i}^{\tns{dyn}}
(\theta) + f_{\tns{wind}}(t_{\tns{wind}}) m_{\tns{ej},i}^{\tns{wind}} (\theta)$,
where $t_{\tns{wstart}}=\unit[20]{ms}$ is the spiral-wave wind onset
\citep{Nedora:2019jhl,Nedora:2020pak} and $t_{\tns{wind}} \in
\unit[[50,200]]{\rm ms}$. The advantage of using such a rescaling procedure is
twofold: i) to align the outcome of simulations with different durations, ii) to
explore the effect of longer winds, whose duration is still comparable to the
simulated one.

% relevant quantities to be computed
For a \ac{BNS} merger for which $\tilde{\theta}$ is the polar angle pointing
toward Earth, and that happened at a time $\unit[t \gg 30]{yr}$ in the past, the
abundance ratio for isotopes $i$ and $j$ of mass numbers $A_i$ and $A_j$ is
\begin{equation}
\frac{Y_i}{Y_j}(\tilde{\theta},t_{\tns{wind}}) =
\frac{A_j}{A_i}
\frac{m_{{\tns{ej}},i}(\tilde{\theta},t_{\tns{wind}})}{m_{{\tns{ej}},j}(\tilde{\theta},t_{\tns{wind}})}
e^{t (1/\tau_j - 1/\tau_i)} \,.
\label{eq:isotopic ratio}
\end{equation}
The fluence of isotope $i$ measured on the Earth, $\mathcal{F}_i$, is related to
the mass of the ejecta and to its radioactivity distance $D_{\tns{rad},i}$ by
\begin{equation}
\mathcal{F}_i = f_{{\tns{dust}},i} \frac{m_{{\tns{ej}},i}^{\tns{iso}}(\tilde{\theta},t_{\tns{wind}})/
\left( A_{i} m_u \right)}{4 \pi D_{{\tns{rad}},i}^2} e^{-t/\tau_i} \,,
\label{eq:fluence}
\end{equation}
where $f_{{\tns{dust}},i}$ is the fraction of $i$ isotopes that form dust,
$m_{{\tns{ej}},i}^{\tns{iso}}$ the isotropized ejecta mass of the isotope $i$
emitted in the direction $\tilde{\theta}$ and $m_u$ the atomic mass unit. For
consistency with \citet{Wallner2021}, we set $f_{{\tns{dust}},\scriptsize\pu} =
f_{{\tns{dust}},\scriptsize\fe} = 0.5$. For $\pu$, we use the fluence value
reported in Table~2 of \citet{Wallner2021}, while the fluence of $\fe$ is
calculated assuming the same fluence over layer incorporation ratio of $\pu$ (also
taken from Table~2 of \citet{Wallner2021}).

% evolution of the ejecta
To gauge the viability of our \ac{BNS} scenario, the radioactivity distance of
different isotopes must be mutually compatible and has to be compared with some
relevant length scales. The first is the fading radius, $R_{\tns{fade}}$. Upon
expulsion, the ejecta expand homologously. Then they enter first a self-similar
Sedov–Taylor expansion phase, and then a snow-plow phase. Finally they dissolve
into the \ac{ISM} upon reaching $R_{\tns{fade}}$ \citep{Montes:2016xkx}. Using
the model outlined in \citet{Beniamini:2017qqy, Bonetti:2019fxj}, assuming
fiducial values for the ejecta mass and speed of \unit[0.04]{$\msun$} and
\unit[0.2]{c}, respectively, and a neutral hydrogen density of
$\unit[0.05]{atoms~cm^{-3}}$ for the Local Bubble \citep{Zucker.etal:2022}, we
estimate $R_{\tns{fade}}\simeq\unit[240]{pc}$. The second length scale is the
radius of the Local Bubble, $R_{\tns{LB}} \simeq \unit[165\pm6]{pc}$
\citep{Zucker.etal:2022}. We also estimate the typical timescale the \ac{BNS}
ejecta would need to expand to such radii. Considering the kilonova remnant
models presented above and the outcome of kilonova remnant simulations
\citep{Montes:2016xkx}, the remnant radius could reach \unit[$\sim$100]{pc}
within \unit[$\sim$1]{Myr}. Finally, we also assume that until fading into the
\ac{ISM} the ejecta do not undergo large-scale mixing, so the angular dependence
of the ejecta is relevant and they enrich the surrounding space in an
anisotropic way. Under this assumption, the isotopic ratios emerging from
the \ac{BNS} event can be directly reflected into those at the Earth's orbit
once accounting for the decay, since $\fe$ and $\pu$ (and all the other live
radioactive isotopes) expand across the Local Bubble with the same history.

\section{Results}
\label{sec:results}

\begin{figure}
  \centering
  \includegraphics[width=\columnwidth]{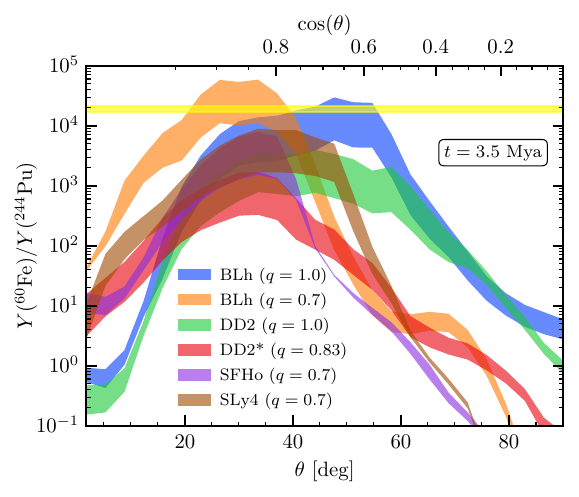}
  \caption{Isotopic ratio between $\fe$ and $\pu$ for all the models listed in
\tabref{tab_1} as a function of the polar angle $\theta$. The bands represent
the variability in the wind duration, $t_{\tns{wind}}\in[50,200]$ ms, with
larger (smaller) ratios related to a longer (shorter) duration. The horizontal
band corresponds to the measured ratio \citep{Wallner2021}.}
    \label{fig:fig_1}
\end{figure}

In \figref{fig_1}, we present $Y_{{~}^{60}{\rm Fe}}/Y_{{~}^{244}{\rm Pu}}$ for a
merger happening $t=3.5$ \ac{Mya}. For the $q=1$ and $q=0.7$ BLh models there
exists a relatively broad range of polar angles at mid-high latitude ($20^\circ
\lesssim \theta \lesssim 60^\circ$) for which the ratio matches the observed
value. The angular intervals are $ 40^\circ \lesssim \theta_1 \lesssim 60^\circ
$ and $ 20^\circ \lesssim \theta_2 \lesssim 40^\circ$ for $q=1$ and $q=0.7$,
respectively. For matter expelled at those latitudes, $\pu$ is mostly produced
in the dynamical ejecta, while the $\fe$ comes from the spiral-wave wind
(see also \figref{fig_4}). In particular, for $20^\circ \lesssim \theta
\lesssim 70^\circ $ the distribution of $\fe$ in the spiral-wave wind ejecta is
rather flat and not very sensitive to the mass ratio. Instead, the $\pu$
distribution decreases moving from the equator to the poles but with a shallower
profile for the unequal mass case. Models other than the BLh ones fail to
reproduce the observed ratio. We notice however that all the simulations share a
similar behavior and some of them still produce a ratio which is not too far
from the observed one. In the case of the DD2 models, the spiral-wave wind is
not rich enough in $\fe$ with respect to the plutonium-rich dynamical ejecta. In
the case of the SFHo and SLy4 models, the amount of $\fe$ is comparable to the
one in the BLh $q=0.7$ case, but the amount of $\pu$ is one order of magnitude
larger. In the rest of the analysis we will focus on the BLh models that match
the measured isotopic ratio.

\begin{figure}
  \centering
  \includegraphics[width=\columnwidth]{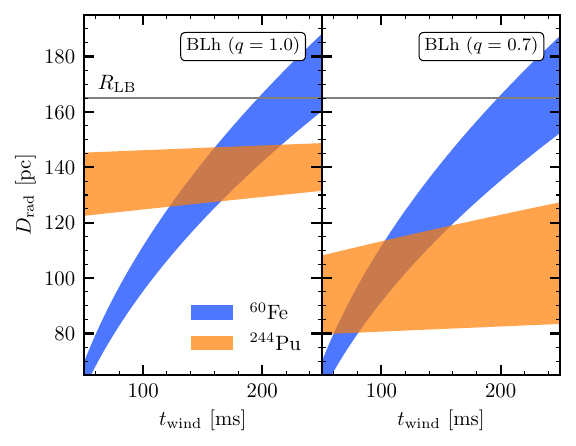}
  \caption{Radioactivity distances as a function of $t_{\tns{wind}}$ for the BLh
models with $q=1.0$ (left) and $q=0.7$ (right). The \ac{BNS} merger is assumed
to occur $t=3.5$ \ac{Mya}. The bands account for different polar angles, \ie
$\theta_1 \in [40^\circ,60^\circ]$ for $q=1$, $\theta_2 \in [20^\circ,40^\circ]$
for $q=0.7$. The horizontal lines mark $R_{\tns{LB}}$ and $R_{\tns{fade}}$.}
    \label{fig:fig_2}
\end{figure}

In \figref{fig_2} we present the radioactivity distance for both $\fe$ and
$\pu$. For each \ac{BNS} model, we consider a variable spiral-wave wind duration
and we span the angular intervals $\theta_1$ and $\theta_2$ presented above. For
both models, there exist a relatively broad $t_{\tns{wind}}$ interval in which
the radioactivity distances of the two isotopes are mutually compatible.
Depending on the model, $D_{\tns{rad}}$ is compatible with an explosion
happening between $\sim$80 and \unit[150]{pc} from Earth. These values are
roughly comparable with the Local Bubble radius. Crucially, they are sufficiently
distant from the Earth to avoid life extinction \citep[$D_{\tns{rad}} \gtrsim
10~{\rm pc}$, see e.g.][]{Perkins:2023iaq} but not too distant for the kilonova
remnant to dissolve before reaching the Earth (\ie $D_{\tns{rad}}\lesssim
R_{\tns{fade}}$).

\begin{figure}
  \centering
  \includegraphics[width=\columnwidth]{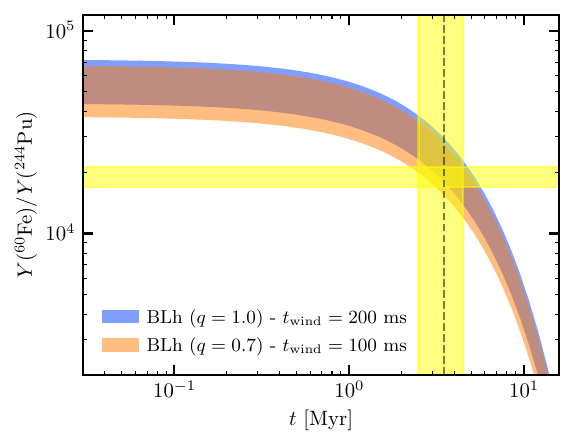}
  \caption{Isotopic ratio between $\fe$ and $\pu$ as a function of time for the
BLh, $q=1.0$ model with $t_{\tns{wind}}=\unit[200]{ms}$ (blue) and for the BLh,
$q=0.7$ model with $t_{\tns{wind}}=\unit[100]{ms}$ (orange). The bands represent
the variability in the polar angle considering the same intervals as in
\figref{fig_2}. The horizontal and vertical bands show instead the measured
ratio \citep{Wallner2021} and a $\unit[\pm1]{Myr}$ uncertainty about the
fiducial value \unit[$t=3.5$]{\ac{Mya}}, respectively.}
    \label{fig:fig_3}
\end{figure}

Given the possible $\unit[\sim 1]{Myr}$ delay between the merger and the arrival
of the ejecta on the Earth, we test the robustness of our results with respect
to the explosion time. In \figref{fig_3} we present the temporal evolution of
$Y_{{~}^{60}{\rm Fe}}/Y_{{~}^{244}{\rm Pu}}$. The results show that our \ac{BNS}
models that can reproduce the observed isotopic ratio and compatible
radioactivity distances can also accommodate a relatively large explosion time
uncertainty ($\unit[\pm1]{Myr}$).

\section{Discussion}
\label{sec:discussion}

\begin{deluxetable}{cccc}
\tablecolumns{4}
\tablecaption{Selected Radioisotope Ratios with Respect to $\pu$ for the BLh
$q=1.0$ and $q=0.7$ \ac{BNS} Models}
\tablehead{
\colhead{Ratio} & \colhead{$\tau$} & \colhead{BLh} & \colhead{BLh}   \\
\colhead{\dots $/\pu$} & (Myr)   & \colhead{$q=1.0$} & \colhead{$q=0.7$}
}
\startdata
$^{93}{\rm Zr}/^{244}{\rm Pu}$ & 2.32 & $8.7 _{-4.5} ^{+6.2} \times 10^{3}$ & $1.2 _{-0.7} ^{+1.0} \times 10^{4}$ \\
$^{107}{\rm Pd}/^{244}{\rm Pu}$ & 9.38 & $2.1 _{-1.2} ^{+1.7} \times 10^{4}$ & $2.8 _{-2.0} ^{+3.3} \times 10^{4}$ \\
$^{129}{\rm I}/^{244}{\rm Pu}$ & 22.65 & $7.6 _{-4.8} ^{+5.9} \times 10^{3}$ & $1.7 _{-1.4} ^{+2.6} \times 10^{4}$ \\
$^{135}{\rm Cs}/^{244}{\rm Pu}$ & 1.92 & $6.7 _{-1.3} ^{+1.6} \times 10^{1}$ & $1.1 _{-0.2} ^{+0.3} \times 10^{2}$ \\
$^{182}{\rm Hf}/^{244}{\rm Pu}$ & 12.84 & $3.2 _{-0.4} ^{+0.4} \times 10^{1}$ & $5.2 _{-1.1} ^{+0.8} \times 10^{1}$ \\
$^{236}{\rm U}/^{244}{\rm Pu}$ & 33.76 & $2.0 _{-0.1} ^{+0.1}$ & $2.0 _{-0.1} ^{+0.1}$ \\
$^{237}{\rm Np}/^{244}{\rm Pu}$ & 3.09 & $5.3 _{-0.2} ^{+0.2} \times 10^{-1}$ & $5.4 _{-0.2} ^{+0.4} \times 10^{-1}$ \\
$^{247}{\rm Cm}/^{244}{\rm Pu}$ & 22.51 & $3.2 _{-0.1} ^{+0.1} \times 10^{-1}$ & $3.0 _{-0.1} ^{+0.1} \times 10^{-1}$ \\
\enddata 
\tablecomments{The intervals span uncertainties in the polar angle and
$t_{\tns{wind}}$. The \ac{BNS} merger is assumed to occur
$\unit[t=3.5]{\ac{Mya}}$.}
\label{tab:tab_2}
\end{deluxetable}

Our results show that the coincident excess of $\fe$ and $\pu$ observed in deep
ocean sediments, dating back to $\unit[\sim3-4]{\ac{Mya}}$, can be explained as
the result of a single \ac{BNS} merger that happened \unit[80-150]{pc}
away from our Solar System.

The difference between our results and the ones presented in several previous
papers \citep[e.g.][]{Fry:2014yqa,Wang2021,Wang2023} can be understood in terms
of some specific features that characterize our \ac{BNS} models. First of all,
the merger remnant must consist in a massive \ac{NS}, not collapsing to a black
hole over a time scale of \unit[100-200]{ms}, to produce a significant
spiral-wave wind ejecta in addition to the dynamical ejecta. The presence of
both these two components is essential, since $\fe$ is synthesized in the
former, while a significant amount of $\pu$ in the latter. Depending on the
\ac{EOS} stiffness and on the colliding \ac{NS} masses, such an outcome is
expected to be relatively frequent and possibly larger than 50\% of the cases
\citep{Margalit:2019dpi}, especially if the recent detections of massive
\acp{NS} will be confirmed \citep{Fonseca:2021wxt,Riley:2021pdl,Romani:2022jhd}.
Moreover, in our analysis we retain information about the angular distribution
of the ejecta and we find that the precise conditions to match the observations
are realized only in matter expelled at mid-high latitudes, \ie for a viewing
angle $30^\circ \lesssim \tilde{\theta} \lesssim 50^\circ$, with an angular
width of $\Delta \theta \approx 20^\circ$. The corresponding solid angle
fraction is $\Delta \Omega/4\pi = 2 \sin{\tilde{\theta}} \sin{(\Delta \theta/2)}
\approx 0.35 \sin{\tilde{\theta}}$, which ranges between 0.18 and 0.27. Despite
not being realized in the majority of cases, the probability of observing a
\ac{BNS} merger in those conditions is not negligible and not even small.

Since our models disfavor viewing angles very close to the poles, the
relativistic jet that could have originated from such an event would not have
hit the Earth due to its small opening angle \citep[$\theta_{\tns{jet}} \lesssim
6^\circ$, see \eg][]{Fong:2015oha,Perkins:2023iaq}. Moreover, the presence of a
relatively broad range of $\theta$ still allows the possibility for the ejecta
to mix, at least over an angular scale $\lesssim \Delta \theta$, due to the
lateral expansion that becomes relevant once the strong blast wave has converted
much of its kinetic energy into thermal energy \citep{Montes:2016xkx}. Large
scale, turbulent mixing is expected to occur only once the expansion speed has
reached the \ac{ISM} sound speed ($\unit[\sim 10]{km~s^{-1}}$) and the kilonova
remnant starts to fade away, on timescales of a few hundreds Myr
\citep{Hotokezaka:2015zea, Beniamini:2020ucb,Kolborg:2023nqu}.

It must be noticed how \ac{BNS} mergers (and, more in general, events in
which \rprocess\ nucleosynthesis occurs) are expected to be rare
\citep{Hotokezaka:2015zea,Abbott2023}, making their nearby occurrence 
in the recent past an exceptional event. Additionally, our analysis does
not rule out the single supernova origin or the two-step model discussed in
previous works, \eg \citet{Wallner2021,Wang2021,Wang2023, Wehmeyer:2023bmg}. The
identification of other relevant isotopic ratios could be the key to
discriminate between the different scenarios, as suggested in
\citet{Wang2021,Wang2023}. To this end, in \tabref{tab_2} we provide the
isotopic ratios with respect to $\pu$ of eight other live radioactive isotopes for
the two representative \ac{BNS} models discussed in \figref{fig_2} and
\figref{fig_3}. Within each model, for ${}^{93}{\tn{Zr}}$, ${}^{107}{\tn{Pd}}$
and ${}^{129}{\tn{I}}$, the values of the isotopic ratios are proportional to
$t_{\tns{wind}}$, while for ${}^{135}{\tn{Cs}}$, ${}^{182}{\tn{Hf}}$ and all the
actinides the dependence on $t_{\tns{wind}}$ is weak or even negligible. Among
the different ratios, the largest values are observed for ${}^{107}{\tn{Pd}}$,
followed by ${}^{93}{\tn{Zr}}$ and ${}^{129}{\tn{I}}$, lower by one order of
magnitude. These trends are different compared with the values presented in
\citet{Wang2021,Wang2023}, for which the largest ratio is always realized for
${}^{129}{\tn{I}}$. For these isotopes, as well as for ${}^{135}{\tn{Cs}}$ and
${}^{182}{\tn{Hf}}$, the values extracted from our models are intermediate
between the larger values obtained by magnetically driven \ac{SN} and the
smaller ones obtained by considering enhanced \rprocess \ac{SN} wind models. The
ratios extracted for the actinides are similar to the ones reported by
\citet{Wang2021,Wang2023}, confirming that they have a low discriminating power.
We also look at the production of $\mn$, that we find to occur only for
very specific thermodynamics conditions ($\ye \gtrsim 0.45$). Our \ac{BNS}
models are not able to reproduce the $\mn$ over $\fe$ ratio of 2:1 in the
interstellar dust predicted by \citet{Korschinek:2020kfp}, due to the very small
amount of ejecta with such high electron fraction (\unit[$\lesssim
10^{-6}$]{$\msun$}). However, more recent \ac{BNS} merger simulations employing
more detailed neutrino transport \citep{Zappa:2022rpd,Espino:2023mda} suggest
the presence of a significant amount of ejecta in the high $\ye$ tail,
especially in the case of long-lived remnants, matching the conditions required
to produce $\mn$.

In our analysis we focused on the coincident $\fe$ and $\pu$ peaks observed in
the youngest deep oceanic crust (\unit[3-4]{\ac{Mya}}), for which the
amount and quality of data are more significant. Due to the paucity of
expected nearby \ac{BNS} mergers, it is implausible that a similar event can
also explain the previous, smaller $\fe$ peak, especially if associated with a
nonnegligible amount of $\pu$. However, several alternative solutions were
previously discussed for the interpretation of the older peak, which apply also
to our scenario, including the fact that the older peak could originate from
outside the Local Bubble \citep[e.g.][]{Schulreich:2023tlp}.

We stress that the single \ac{BNS} event scenario can explain the observational
data without any need for tuning. Indeed the only free parameter in our model is
the spiral-wave wind duration, and we conservatively vary it over a range
comparable to the duration of our simulations. To better address the viability
of the kilonova scenario, more realistic models would be necessary. But, if
confirmed, our analysis shows that the merger of a \ac{BNS} system could have
happened in the Solar neighborhood no earlier than $\unit[\sim4]{\ac{Mya}}$.

\begin{acknowledgments}  
The authors thank B. Wehmeyer for helpful discussions, the Computational Relativity
(CORE) collaboration for providing simulation results, and the European Centre
for Theoretical Studies in Nuclear Physics and Related Areas (ECT$^*$) for hosting
the MICRA2023 workshop, where this work was conceived. They also thank the INFN
for the usage of computing and storage resources through the tullio cluster in
Turin. This work has been supported by STRONG-2020 ``The strong interaction at
the frontier of knowledge: fundamental research and applications'' which received
funding from the European Union’s Horizon 2020 research and innovation programme
under grant agreement No 824093. The work of AP is partially funded by the European
Union under NextGenerationEU. PRIN 2022 Prot. n. 2022KX2Z3B. FMG acknowledges funding
from the Fondazione CARITRO, program Bando post-doc 2021, project number 11745.
\end{acknowledgments}

\appendix

\section{Nucleosynthesis calculation in \ac{BNS} ejecta}
\label{sec:appendix}

\begin{figure}
    \centering
    \includegraphics[width=\columnwidth]{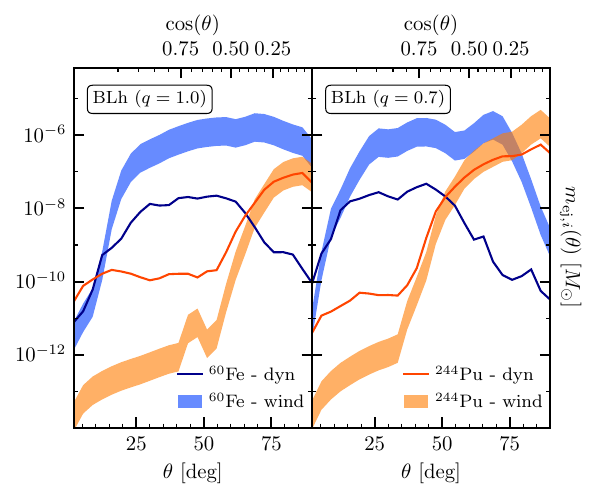}
    \caption{$\fe$ and $\pu$ masses ejected in BLh models with $q=1.0$
(left) and $q=0.7$ (right) as a function of the polar angle $\theta$. The
contribution from the dynamical and spiral-wave wind ejecta are shown
separately. The band represents the variability in the wind duration,
$t_{\tns{wind}}\in[50,200]$ ms, with larger (smaller) isotopic masses related to
a longer (shorter) duration.}
    \label{fig:fig_4}
\end{figure}

% nucleosynthesis calculation
In order to compute the total isotopic masses expelled in each \ac{BNS}
simulation, we exploit the outcome of an extensive set of nuclear network
calculations performed on a grid of 11700 Lagrangian tracer particles. The grid
spans broad ranges in the space of initial thermodynamic conditions parametrized
by ($s$, $\ye$, $\tau_{\tns{exp}}$), and is identical to the one used in
\citet{Perego:2020evn}. We employ the \texttt{SkyNet} nuclear network
\citep{Lippuner:2017tyn} to evolve in time the number abundances of a wide set
of nuclear species, including $\fe$ and $\pu$, depending on the specific
Lagrangian particle's history. The initial nuclear distribution follows from
\ac{NSE} conditions that are determined by the ($s$,$\ye$) values of the
trajectory, at a temperature fixed to \unit[8]{GK} (high enough to guarantee the
validity of the \ac{NSE} assumption). The composition is then evolved along the
analytic density profile used in \citet{Lippuner:2015gwa}, parametrized by
$\tau_{\tns{exp}}$. The network is run until the final time of $\sim31.7$ years,
using the same input nuclear physics as in \citet{Perego:2020evn}. 
To obtain the polar angle dependent yields produced in the merger event we
compute the convolution of the mass histograms of the ejecta (as described in
the main text and directly extracted from the simulations) with the abundances
obtained on the $(s,\ye,\tau_{\tns{exp}})$ space by the nuclear network evolution.
We proceed separately for the dynamical and spiral-wave wind ejecta components.
The total isotopic masses in the ejecta are then obtained by
summing the two contributions, after having rescaled the spiral-wave wind one
for the corresponding $t_{\rm wind}$ as discussed in \secref{methods}.
\figref{fig_4} shows the amount of $\fe$ and $\pu$ produced at 30 years
after merger in the two representative \ac{BNS} models of \figref{fig_2} and
\figref{fig_3}. In the proposed \ac{BNS} scenario, $\fe$ is synthesized over a
wide angular range and mainly in the spiral-wave wind ejecta. These ejecta are
characterized by a relatively high $\ye$, an effect of the prolonged neutrino
irradiation from the central long-lived remnant. A similar behavior is observed
for $^{93}{\rm Zr}$ and $^{107}{\rm Pd}$, listed in \tabref{tab_2}. The
production of $\pu$ instead, representative of the heavier isotopes in the
table, peaks in the equatorial region ($\theta \gtrsim 60^\circ$) and originates
from the fraction of cold, neutron-rich matter present in the two ejecta
components.

\bibliographystyle{aasjournal}

\begin{thebibliography}{}
\expandafter\ifx\csname natexlab\endcsname\relax\def\natexlab#1{#1}\fi
\providecommand{\url}[1]{\href{#1}{#1}}
\providecommand{\dodoi}[1]{doi:~\href{http://doi.org/#1}{\nolinkurl{#1}}}
\providecommand{\doeprint}[1]{\href{http://ascl.net/#1}{\nolinkurl{http://ascl.net/#1}}}
\providecommand{\doarXiv}[1]{\href{https://arxiv.org/abs/#1}{\nolinkurl{https://arxiv.org/abs/#1}}}

\bibitem[{Abbott {et~al.}(2019)}]{Abbott:2018wiz}
Abbott, B.~P., {et~al.} 2019, Phys. Rev., X9, 011001,
  \dodoi{10.1103/PhysRevX.9.011001}

\bibitem[{Abbott {et~al.}(2023)}]{Abbott2023}
Abbott, R., {et~al.} 2023, Phys. Rev. X, 13, 011048,
  \dodoi{10.1103/PhysRevX.13.011048}

\bibitem[{Bartos \& Marka(2019)}]{Bartos:2019twj}
Bartos, I., \& Marka, S. 2019, Astrophys. J., 881, L4,
  \dodoi{10.3847/2041-8213/ab3215}

\bibitem[{Beniamini {et~al.}(2018)Beniamini, Dvorkin, \&
  Silk}]{Beniamini:2017qqy}
Beniamini, P., Dvorkin, I., \& Silk, J. 2018, Mon. Not. Roy. Astron. Soc., 478,
  1994, \dodoi{10.1093/mnras/sty1035}

\bibitem[{Beniamini \& Hotokezaka(2020)}]{Beniamini:2020ucb}
Beniamini, P., \& Hotokezaka, K. 2020, Mon. Not. Roy. Astron. Soc., 496, 1891,
  \dodoi{10.1093/mnras/staa1690}

\bibitem[{Bernuzzi {et~al.}(2020)}]{Bernuzzi:2020txg}
Bernuzzi, S., {et~al.} 2020, Mon. Not. Roy. Astron. Soc.,
  \dodoi{10.1093/mnras/staa1860}

\bibitem[{Bonetti {et~al.}(2019)Bonetti, Perego, Dotti, \&
  Cescutti}]{Bonetti:2019fxj}
Bonetti, M., Perego, A., Dotti, M., \& Cescutti, G. 2019, Mon. Not. Roy.
  Astron. Soc., 490, 296, \dodoi{10.1093/mnras/stz2554}

\bibitem[{Burbidge {et~al.}(1957)Burbidge, Burbidge, Fowler, \&
  Hoyle}]{Burbidge:1957vc}
Burbidge, M.~E., Burbidge, G.~R., Fowler, W.~A., \& Hoyle, F. 1957, Rev. Mod.
  Phys., 29, 547, \dodoi{10.1103/RevModPhys.29.547}

\bibitem[{Cowan {et~al.}(2021)Cowan, Sneden, Lawler, Aprahamian, Wiescher,
  Langanke, Mart\'\i{}nez-Pinedo, \& Thielemann}]{Cowan:2019pkx}
Cowan, J.~J., Sneden, C., Lawler, J.~E., {et~al.} 2021, Rev. Mod. Phys., 93,
  15002, \dodoi{10.1103/RevModPhys.93.015002}

\bibitem[{Côté {et~al.}(2019)}]{Cote:2018qku}
Côté, B., {et~al.} 2019, Astrophys. J., 875, 106,
  \dodoi{10.3847/1538-4357/ab10db}

\bibitem[{Douchin \& Haensel(2001)}]{Douchin:2001sv}
Douchin, F., \& Haensel, P. 2001, Astron. Astrophys., 380, 151,
  \dodoi{10.1051/0004-6361:20011402}

\bibitem[{Ellis {et~al.}(1996)Ellis, Fields, \& Schramm}]{Ellis:1995qb}
Ellis, J.~R., Fields, B.~D., \& Schramm, D.~N. 1996, Astrophys. J., 470, 1227,
  \dodoi{10.1086/177945}

\bibitem[{Espino {et~al.}(2023)Espino, Radice, Zappa, Gamba, \&
  Bernuzzi}]{Espino:2023mda}
Espino, P.~L., Radice, D., Zappa, F., Gamba, R., \& Bernuzzi, S. 2023.
\newblock \doarXiv{2311.12923}

\bibitem[{Fields {et~al.}(2005)Fields, Hochmuth, \& Ellis}]{Fields:2004iq}
Fields, B.~D., Hochmuth, K.~A., \& Ellis, J.~R. 2005, Astrophys. J., 621, 902,
  \dodoi{10.1086/427797}

\bibitem[{Fong {et~al.}(2015)Fong, Berger, Margutti, \&
  Zauderer}]{Fong:2015oha}
Fong, W.-f., Berger, E., Margutti, R., \& Zauderer, B.~A. 2015, Astrophys. J.,
  815, 102, \dodoi{10.1088/0004-637X/815/2/102}

\bibitem[{Fonseca {et~al.}(2021)}]{Fonseca:2021wxt}
Fonseca, E., {et~al.} 2021, Astrophys. J. Lett., 915, L12,
  \dodoi{10.3847/2041-8213/ac03b8}

\bibitem[{Foucart {et~al.}(2016)Foucart, Haas, Duez, O'Connor, Ott, Roberts,
  Kidder, Lippuner, Pfeiffer, \& Scheel}]{Foucart:2015gaa}
Foucart, F., Haas, R., Duez, M.~D., {et~al.} 2016, Phys. Rev., D93, 044019,
  \dodoi{10.1103/PhysRevD.93.044019}

\bibitem[{Fry {et~al.}(2015)Fry, Fields, \& Ellis}]{Fry:2014yqa}
Fry, B.~J., Fields, B.~D., \& Ellis, J.~R. 2015, Astrophys. J., 800, 71,
  \dodoi{10.1088/0004-637X/800/1/71}

\bibitem[{Gonzalez {et~al.}(2023)}]{Gonzalez:2022mgo}
Gonzalez, A., {et~al.} 2023, Class. Quant. Grav., 40, 085011,
  \dodoi{10.1088/1361-6382/acc231}

\bibitem[{Hempel \& Schaffner-Bielich(2010)}]{Hempel:2009mc}
Hempel, M., \& Schaffner-Bielich, J. 2010, Nucl. Phys., A837, 210,
  \dodoi{10.1016/j.nuclphysa.2010.02.010}

\bibitem[{Hotokezaka {et~al.}(2015)Hotokezaka, Piran, \&
  Paul}]{Hotokezaka:2015zea}
Hotokezaka, K., Piran, T., \& Paul, M. 2015, Nature Phys., 11, 1042,
  \dodoi{10.1038/nphys3574}

\bibitem[{Kasen {et~al.}(2017)Kasen, Metzger, Barnes, Quataert, \&
  Ramirez-Ruiz}]{Kasen:2017sxr}
Kasen, D., Metzger, B., Barnes, J., Quataert, E., \& Ramirez-Ruiz, E. 2017,
  Nature, \dodoi{10.1038/nature24453}

\bibitem[{Kolborg {et~al.}(2023)Kolborg, Ramirez-Ruiz, Martizzi, Macias, \&
  Soares-Furtado}]{Kolborg:2023nqu}
Kolborg, A.~N., Ramirez-Ruiz, E., Martizzi, D., Macias, P., \& Soares-Furtado,
  M. 2023, Astrophys. J., 949, 100, \dodoi{10.3847/1538-4357/acca80}

\bibitem[{{Korschinek} {et~al.}(2020){Korschinek}, {Faestermann}, {Poutivtsev},
  {Arazi}, {Knie}, {Rugel}, \& {Wallner}}]{Korschinek:2020kfp}
{Korschinek}, G., {Faestermann}, T., {Poutivtsev}, M., {et~al.} 2020, \prl,
  125, 031101, \dodoi{10.1103/PhysRevLett.125.031101}

\bibitem[{Lippuner \& Roberts(2015)}]{Lippuner:2015gwa}
Lippuner, J., \& Roberts, L.~F. 2015, Astrophys. J., 815, 82,
  \dodoi{10.1088/0004-637X/815/2/82}

\bibitem[{Lippuner \& Roberts(2017)}]{Lippuner:2017tyn}
---. 2017, Astrophys. J. Suppl., 233, 18, \dodoi{10.3847/1538-4365/aa94cb}

\bibitem[{Logoteta {et~al.}(2021)Logoteta, Perego, \&
  Bombaci}]{Logoteta:2020yxf}
Logoteta, D., Perego, A., \& Bombaci, I. 2021, Astron. Astrophys., 646, A55,
  \dodoi{10.1051/0004-6361/202039457}

\bibitem[{Macias \& Ramirez-Ruiz(2019)}]{Macias:2019oxw}
Macias, P., \& Ramirez-Ruiz, E. 2019, Astrophys. J. Lett., 877, L24,
  \dodoi{10.3847/2041-8213/ab2049}

\bibitem[{Margalit \& Metzger(2019)}]{Margalit:2019dpi}
Margalit, B., \& Metzger, B.~D. 2019, Astrophys. J. Lett., 880, L15,
  \dodoi{10.3847/2041-8213/ab2ae2}

\bibitem[{Montes {et~al.}(2016)Montes, Ramirez-Ruiz, Naiman, Shen, \&
  Lee}]{Montes:2016xkx}
Montes, G., Ramirez-Ruiz, E., Naiman, J., Shen, S., \& Lee, W.~H. 2016,
  Astrophys. J., 830, 12, \dodoi{10.3847/0004-637X/830/1/12}

\bibitem[{M{\"o}sta {et~al.}(2018)M{\"o}sta, Roberts, Halevi, Ott, Lippuner,
  Haas, \& Schnetter}]{Mosta:2017geb}
M{\"o}sta, P., Roberts, L.~F., Halevi, G., {et~al.} 2018, Astrophys. J., 864,
  171, \dodoi{10.3847/1538-4357/aad6ec}

\bibitem[{Nedora {et~al.}(2019)Nedora, Bernuzzi, Radice, Perego, Endrizzi, \&
  Ortiz}]{Nedora:2019jhl}
Nedora, V., Bernuzzi, S., Radice, D., {et~al.} 2019, Astrophys. J., 886, L30,
  \dodoi{10.3847/2041-8213/ab5794}

\bibitem[{Nedora {et~al.}(2021)Nedora, Bernuzzi, Radice, Daszuta, Endrizzi,
  Perego, Prakash, Safarzadeh, Schianchi, \& Logoteta}]{Nedora:2020pak}
---. 2021, Astrophys. J., 906, 98, \dodoi{10.3847/1538-4357/abc9be}

\bibitem[{Perego {et~al.}(2017)Perego, Radice, \& Bernuzzi}]{Perego:2017wtu}
Perego, A., Radice, D., \& Bernuzzi, S. 2017, Astrophys. J., 850, L37,
  \dodoi{10.3847/2041-8213/aa9ab9}

\bibitem[{Perego {et~al.}(2022)}]{Perego:2020evn}
Perego, A., {et~al.} 2022, Astrophys. J., 925, 22,
  \dodoi{10.3847/1538-4357/ac3751}

\bibitem[{Perkins {et~al.}(2024)Perkins, Ellis, Fields, Hartmann, Liu,
  McLaughlin, Surman, \& Wang}]{Perkins:2023iaq}
Perkins, H. M.~L., Ellis, J., Fields, B.~D., {et~al.} 2024, Astrophys. J., 961,
  170, \dodoi{10.3847/1538-4357/ad12b7}

\bibitem[{Pian {et~al.}(2017)}]{Pian:2017gtc}
Pian, E., {et~al.} 2017, Nature, \dodoi{10.1038/nature24298}

\bibitem[{Radice(2017)}]{Radice:2017zta}
Radice, D. 2017, Astrophys. J., 838, L2, \dodoi{10.3847/2041-8213/aa6483}

\bibitem[{Radice(2020)}]{Radice:2020ids}
---. 2020, Symmetry, 12, 1249, \dodoi{10.3390/sym12081249}

\bibitem[{Radice {et~al.}(2016)Radice, Galeazzi, Lippuner, Roberts, Ott, \&
  Rezzolla}]{Radice:2016dwd}
Radice, D., Galeazzi, F., Lippuner, J., {et~al.} 2016, Mon. Not. Roy. Astron.
  Soc., 460, 3255, \dodoi{10.1093/mnras/stw1227}

\bibitem[{Radice {et~al.}(2018)Radice, Perego, Hotokezaka, Fromm, Bernuzzi, \&
  Roberts}]{Radice:2018pdn}
Radice, D., Perego, A., Hotokezaka, K., {et~al.} 2018, Astrophys. J., 869, 130,
  \dodoi{10.3847/1538-4357/aaf054}

\bibitem[{Radice \& Rezzolla(2012)}]{Radice:2012cu}
Radice, D., \& Rezzolla, L. 2012, Astron. Astrophys., 547, A26,
  \dodoi{10.1051/0004-6361/201219735}

\bibitem[{Radice {et~al.}(2014{\natexlab{a}})Radice, Rezzolla, \&
  Galeazzi}]{Radice:2013xpa}
Radice, D., Rezzolla, L., \& Galeazzi, F. 2014{\natexlab{a}},
  Class.Quant.Grav., 31, 075012, \dodoi{10.1088/0264-9381/31/7/075012}

\bibitem[{Radice {et~al.}(2014{\natexlab{b}})Radice, Rezzolla, \&
  Galeazzi}]{Radice:2013hxh}
---. 2014{\natexlab{b}}, Mon.Not.Roy.Astron.Soc., 437, L46,
  \dodoi{10.1093/mnrasl/slt137}

\bibitem[{Riley {et~al.}(2021)}]{Riley:2021pdl}
Riley, T.~E., {et~al.} 2021, Astrophys. J. Lett., 918, L27,
  \dodoi{10.3847/2041-8213/ac0a81}

\bibitem[{Romani {et~al.}(2022)Romani, Kandel, Filippenko, Brink, \&
  Zheng}]{Romani:2022jhd}
Romani, R.~W., Kandel, D., Filippenko, A.~V., Brink, T.~G., \& Zheng, W. 2022,
  Astrophys. J. Lett., 934, L17, \dodoi{10.3847/2041-8213/ac8007}

\bibitem[{Schneider {et~al.}(2017)Schneider, Roberts, \&
  Ott}]{daSilvaSchneider:2017jpg}
Schneider, A.~S., Roberts, L.~F., \& Ott, C.~D. 2017, Phys. Rev., C96, 065802,
  \dodoi{10.1103/PhysRevC.96.065802}

\bibitem[{Schulreich {et~al.}(2017)Schulreich, Breitschwerdt, Feige, \&
  Dettbarn}]{Schulreich:2017dyt}
Schulreich, M.~M., Breitschwerdt, D., Feige, J., \& Dettbarn, C. 2017, Astron.
  Astrophys., 604, A81, \dodoi{10.1051/0004-6361/201629837}

\bibitem[{Schulreich {et~al.}(2023)Schulreich, Feige, \&
  Breitschwerdt}]{Schulreich:2023tlp}
Schulreich, M.~M., Feige, J., \& Breitschwerdt, D. 2023, Astron. Astrophys.,
  680, A39, \dodoi{10.1051/0004-6361/202347532}

\bibitem[{Sekiguchi {et~al.}(2015)Sekiguchi, Kiuchi, Kyutoku, \&
  Shibata}]{Sekiguchi:2015dma}
Sekiguchi, Y., Kiuchi, K., Kyutoku, K., \& Shibata, M. 2015, Phys.Rev., D91,
  064059, \dodoi{10.1103/PhysRevD.91.064059}

\bibitem[{Siegel {et~al.}(2019)Siegel, Barnes, \& Metzger}]{Siegel:2018zxq}
Siegel, D.~M., Barnes, J., \& Metzger, B.~D. 2019, Nature, 569, 241,
  \dodoi{10.1038/s41586-019-1136-0}

\bibitem[{Smartt {et~al.}(2017)}]{Smartt:2017fuw}
Smartt, S.~J., {et~al.} 2017, Nature, \dodoi{10.1038/nature24303}

\bibitem[{Steiner {et~al.}(2013)Steiner, Hempel, \& Fischer}]{Steiner:2012rk}
Steiner, A.~W., Hempel, M., \& Fischer, T. 2013, Astrophys. J., 774, 17,
  \dodoi{10.1088/0004-637X/774/1/17}

\bibitem[{Typel {et~al.}(2010)Typel, Ropke, Klahn, Blaschke, \&
  Wolter}]{Typel:2009sy}
Typel, S., Ropke, G., Klahn, T., Blaschke, D., \& Wolter, H.~H. 2010, Phys.
  Rev., C81, 015803, \dodoi{10.1103/PhysRevC.81.015803}

\bibitem[{{Wallner} {et~al.}(2016){Wallner}, {Feige}, {Kinoshita}, {Paul},
  {Fifield}, {Golser}, {Honda}, {Linnemann}, {Matsuzaki}, {Merchel}, {Rugel},
  {Tims}, {Steier}, {Yamagata}, \& {Winkler}}]{Wallner.etal:2016}
{Wallner}, A., {Feige}, J., {Kinoshita}, N., {et~al.} 2016, \nat, 532, 69,
  \dodoi{10.1038/nature17196}

\bibitem[{Wallner {et~al.}(2021)Wallner, Froehlich, Hotchkis, Kinoshita, Paul,
  Martschini, Pavetich, Tims, Kivel, Schumann, Honda, Matsuzaki, \&
  Yamagata}]{Wallner2021}
Wallner, A., Froehlich, M.~B., Hotchkis, M. A.~C., {et~al.} 2021, Science, 372,
  742, \dodoi{10.1126/science.aax3972}

\bibitem[{Wang {et~al.}(2021)Wang, Clark, Ellis, Ertel, Fields, Fry, Liu,
  Miller, \& Surman}]{Wang2021}
Wang, X., Clark, A.~M., Ellis, J., {et~al.} 2021, The Astrophysical Journal,
  923, 219, \dodoi{10.3847/1538-4357/ac2d90}

\bibitem[{Wang {et~al.}(2023)Wang, Clark, Ellis, Ertel, Fields, Fry, Liu,
  Miller, \& Surman}]{Wang2023}
---. 2023, The Astrophysical Journal, 948, 113,
  \dodoi{10.3847/1538-4357/acbeaa}

\bibitem[{Wehmeyer {et~al.}(2023)Wehmeyer, L\'opez, C\^ot\'e, Pet\H{o},
  Kobayashi, \& Lugaro}]{Wehmeyer:2023bmg}
Wehmeyer, B., L\'opez, A.~Y., C\^ot\'e, B., {et~al.} 2023, Astrophys. J., 944,
  121, \dodoi{10.3847/1538-4357/acafec}

\bibitem[{Winteler {et~al.}(2012)Winteler, Kaeppeli, Perego, Arcones, Vasset,
  Nishimura, Liebendoerfer, \& Thielemann}]{Winteler:2012hu}
Winteler, C., Kaeppeli, R., Perego, A., {et~al.} 2012, Astrophys. J. Lett.,
  750, L22, \dodoi{10.1088/2041-8205/750/1/L22}

\bibitem[{Zappa {et~al.}(2023)Zappa, Bernuzzi, Radice, \&
  Perego}]{Zappa:2022rpd}
Zappa, F., Bernuzzi, S., Radice, D., \& Perego, A. 2023, Mon. Not. Roy. Astron.
  Soc., 520, 1481, \dodoi{10.1093/mnras/stad107}

\bibitem[{{Zucker} {et~al.}(2022){Zucker}, {Goodman}, {Alves}, {Bialy},
  {Foley}, {Speagle}, {Gro{\^I}{\texttwosuperior}schedl}, {Finkbeiner},
  {Burkert}, {Khimey}, \& {Swiggum}}]{Zucker.etal:2022}
{Zucker}, C., {Goodman}, A.~A., {Alves}, J., {et~al.} 2022, \nat, 601, 334,
  \dodoi{10.1038/s41586-021-04286-5}

\end{thebibliography}

\end{document}